\documentclass[a4paper]{jpconf}
\usepackage{graphicx}
\usepackage[font=small,labelfont=bf,tableposition=top]{caption}
\usepackage{lineno}
\DeclareCaptionLabelFormat{andtable}{#1~#2  \&  \tablename~\thetable}
\usepackage{cite}

\begin{document}
\title{Correlation of heavy-flavour hadron production and charged-particle multiplicity in pp collisions measured by ALICE}

\author{Joyful Elma Mdhluli for the ALICE Collaboration}
\address{University of the Witwatersrand, 1 Jan Smuts, Johannesburg, 2000, South Africa.}
\ead{joyful.elma.mdhluli@cern.ch}

\begin{abstract}
Heavy-flavour (HF) quarks, i.e. charm and beauty, are produced in the early stages of ultra-relativistic collisions via hard scattering processes. The measurement of heavy flavour as a function of charged-particle multiplicity not only provides information on how the production mechanisms are influenced by the event activity, but can also be considered as a tool to test the role of multiple parton interactions. Furthermore, charged-particle multiplicity studies are essential for reference measurements, to tune theoretical models and to provide information on the global characteristics of events. In ALICE, charged-particle multiplicity and heavy-flavour production in the hadronic and electronic decay channels are measured at central rapidity while HF-decay muons and $\rm J/\psi \rightarrow \mu^{+}\mu^{-}$ are measured at forward rapidity. In this contribution, results will be presented on relative $\rm J/\psi$, average D-mesons and leptons from HF hadron decay yields as a function of charged-particle multiplicity in proton-proton (pp) collisions at $\sqrt{s}$ = 5.02, 7, 8 and 13 TeV. These results will also be compared to theoretical model calculations. 
\end{abstract}

\section{Introduction}
Heavy-flavour quarks, i.e. charm and beauty, are produced via the initial hard parton-parton scattering processes in the early stages of ultra-relativistic collisions. Charm and beauty quarks are produced abundantly at the energies accessed at the  Large Hadron Collider (LHC) in pp collisions. Heavy-flavour decay muon measurements with A Large Ion Collider Experiment (ALICE) at forward rapidity ($2.5 < y < 4$) were published for the production cross section as a function of transverse momentum ($p_{\rm T}$) and rapidity ($y$) in pp collisions at $\sqrt{s}$ = 2.76, 5.02 and 7 TeV  \cite{run1, run2, run1.1} . Predictions of models based on pQCD calculations, such as FONLL and NLLO \cite{FONLL, NNLO}, describe the data within the experimental and theoretical uncertainties. An energy dependence is observed where the production cross-section of beauty and charm quarks increases with center-of-mass energy. Measurements of heavy-flavour hadron production cross sections in pp collisions are important for testing theoretical predictions.\\ 

The measurement of the heavy-flavour production as a function of the charged-particle multiplicity provides tools to investigate the role of multiple parton interaction (MPI)  in the production of particles and the role of the color reconnection in the hadronization mechanisms process, in particular in hadron rich environments like those of proton and nucleus collisions at the LHC. At LHC energies, MPIs give a major contribution to particle production, and theoretical models must  account for MPI in order to describe, at least qualitatively, the dependence of heavy-flavour production on charged-particle multiplicity. At high charged-particle multiplicities in pp and proton$-$nucleus (p$-$A) collisions intriguing effects, such as strangeness enhancement \cite{strangeness} and a positive elliptic flow were found that are typically observed in nucleus$-$nucleus (A$-$A) collisions where they are interpreted  as a signature of collective motion of an expanding medium \cite{collectivity}. Positive elliptic-flow coefficients were measured also for heavy-flavour signals \cite{collectivity1}. Multiplicity-dependent measurements allow the study of the interplay between soft and hard particle production mechanisms. Results from production measurements of D mesons, $\rm J/\psi$ and leptons from heavy-flavour decay at mid- and forward rapidity as a function of charged-particle multiplicity are presented in this contribution. 

\section{ALICE apparatus}
The ALICE apparatus at the LHC was designed to study strongly-interacting matter at extreme temperature and energy density. The detector consists of 19 subdetectors and is described in detail in reference \cite{Aamodt}. The central barrel, which covers the pseudorapidity interval $|\eta| < 0.9$, is responsible for measuring hadrons, electrons and photons, while; the forward muon spectrometer, which covers the pseudorapidity interval $-4 < \eta < -2.5$, is dedicated to measure muons. The main central barrel subdetectors utilized for heavy-flavour production measurements are the Inner Tracking System (ITS), Time Projection Chamber (TPC), Time-Of-Flight Detector (TOF), ElectroMagnetic Calorimeter (EMCal). The charged-particle multiplicity is measured with the V0 detector consisting of two scintillator arrays (V0A and V0C) covering the ranges $2.8 < \eta < 5.1$ and $-3.7 < \eta < -1.7$, and with the two innermost layers of the ITS, the silicon pixel detector (SPD). 

\section{Measurements}
\subsection{Heavy-flavour production} 
In ALICE, the D-meson production is measured via the reconstruction of hadronic decays at mid-rapidity. A detailed discussion on the measurement of the D-meson yield is given in reference \cite{dmeson1}. Quarkonium production is measured via the electronic decay channel ($\textrm{J}/\psi \rightarrow \rm e^{-}e^{+}$) at midrapidity and via the muonic decay channel ($\textrm{J}/\psi \rightarrow \rm \mu^{+}\mu^{-}$) at forward rapidity. A detailed discussion on the measurement of the multiplicity dependence of $\textrm{J}/\psi$ production is given in reference \cite{jpsi}. The measurements of the multiplicity dependence of HF decay electrons (HFe) and of HF decay muons (HFm), $\textrm{c,b} \rightarrow \rm e/\mu +X$, at mid- and forward rapidity, respectively, are reported in references \cite{run2} and \cite{HFe}.

\subsection{Multiplicity selection}
Two multiplicity estimators were used to define the multiplicity intervals: the SPD estimator and the V0M estimator. The SPD estimator is based on measurements at midrapidity and it is determined by the number of reconstructed "tracklets" in the two innermost layers of the ITS (SPD). The V0M estimator is extracted from measurements at forward rapidity and it is determined by obtaining the total percentiles of the V0 amplitude distributions of the V0A and V0C detectors.

\section{Results}
\begin{figure}[htb!]
\centering
\begin{minipage}{16pc}
\includegraphics[width=16pc]{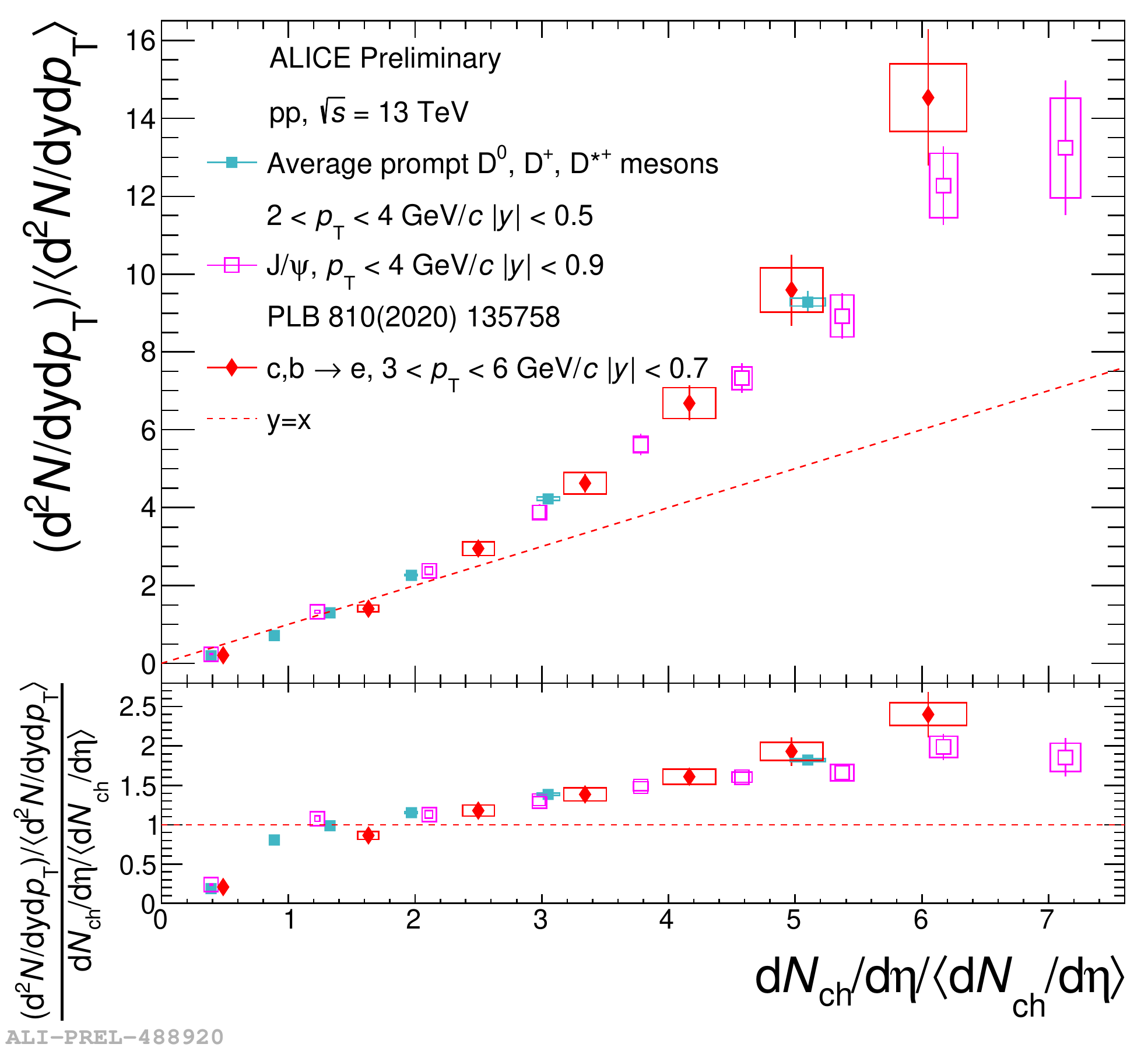}
\caption{\label{mid}Comparison of the average D-meson, $J/\psi$, and HFe self-normalised yields vs. relative charged-particle multiplicity at central rapidity in pp collisions at $\sqrt{s}$ = 13 TeV \cite{dmeson}.}
\end{minipage}\hspace{1pc}%
\begin{minipage}{18pc}
\centering
\includegraphics[width=18pc]{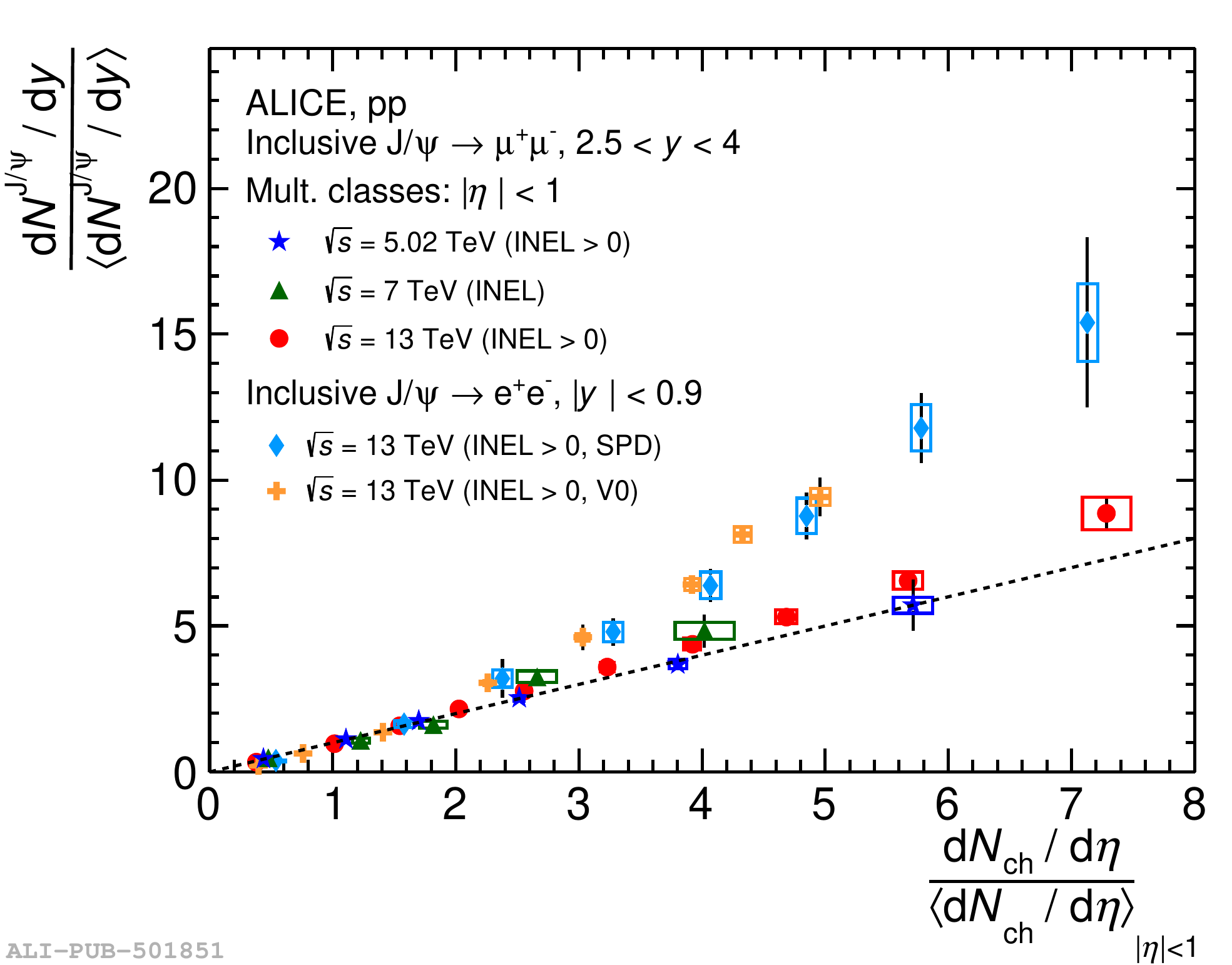}
\caption{\label{forward}Relative $J/\psi$ yields as a function of the relative charged-particle multiplicity measured at mid- and forward rapidity in pp collisions at $\sqrt{s} = 5.02$ (INEL$>0$), 7 (INEL) and 13 (INEL$>0$) TeV \cite{jpsi}.}
\end{minipage} 
\end{figure}

A comparison of the average D mesons, $\rm J/\psi$, and heavy-flavour electron self-normalised yields (yield measured in a multiplicity class divided by that measured without multiplicity selection) vs. relative charged-particle multiplicity at central rapidity in pp collisions at $\sqrt{s}$ = 13 TeV is shown in Fig. \ref{mid} \cite{dmeson}. The D-meson yields are shown for $2 < p_{\rm T} < 4$ GeV/$c$, at $|y| <$ 0.5, $\rm J/\psi$ yields are shown for $p_{\rm T} < 4$ GeV/$c$, at $|y| < $0.9, and HFe yields are shown for $3 < p_{\rm T} < 6$ GeV/$c$, at $|y| <$ 0.7. The self-normalised yields are presented in the top panel with their statistical (vertical bars) and systematic (boxes) uncertainties. The bottom panel shows the comparison of the double ratios between the self-normalised yields and the relative charged-particle multiplicity, as a function of relative charged-particle multiplicity. A similar trend at midrapidity for different particle species (D mesons, HFe, $\textrm{J}/\psi$) is observed, suggesting a common origin for the increasing trend.\\

The relative $\textrm{J}/\psi$ yields as a function of the relative charged-particle multiplicity measured at mid- and forward rapidity in pp collisions at $\sqrt{s} = 5.02$, 7 and 13 TeV are shown in Fig. \ref{forward}. The measurements differ for the physical event class they refer to: inelastic events (INEL) for the measurement at $\sqrt{s} = 7$ TeV, inelastic events with at least one charged particle in $|\eta| < 1$ for the measurements at $\sqrt{s} = 5.02$ and 13 TeV. The vertical bars and the boxes represent the statistical and the systematic uncertainties, respectively. A close-to-linear trend is observed for measurements at forward rapidity at various collision energies, suggesting that there is no energy dependence of the relative $\textrm{J}/\psi$ production in the same relative final-state multiplicity domain \cite{jpsi}. In addition, a comparison between the forward and midrapidity measurements at $\sqrt{s}$ = 13 TeV is also reported. A faster-than-linear increase as a function of charged-particle multiplicity is observed in the midrapidity measurements. The results using midrapidity multiplicity selection based on the SPD detector and forward rapidity multiplicity selection based on the V0 detectors are also shown in Fig. \ref{forward} and found to be compatible within uncertainties. This suggests that, for $\rm J/\psi$ mesons, the difference in trends between the forward and midrapidity measurements is not due to possible auto-correlation bias that arises from the multiplicity selection \cite{jpsi}.\\

\begin{figure}[htb!]
\centering
\begin{minipage}{18pc}
\includegraphics[width=19pc]{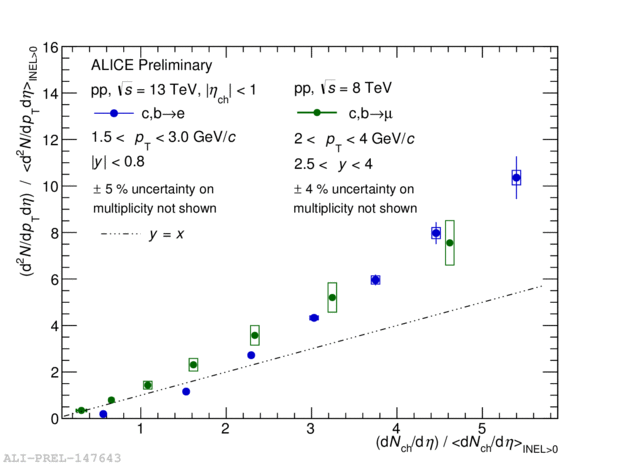}
\caption{\label{data}Comparison of the self-normalized yield vs. charged-particle multiplicity for forward-$y$ HFm and midrapidity HFe decay \cite{HFe}.}
\end{minipage}\hspace{1pc}%
\begin{minipage}{18pc}
\centering
\includegraphics[width=19pc]{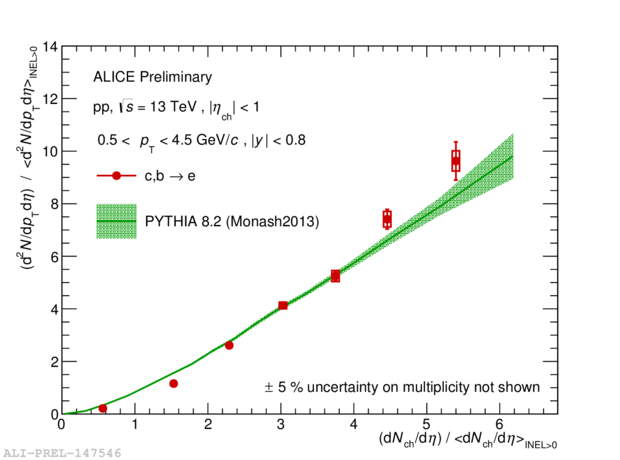}
\caption{\label{model}Comparison of the self-normalized yield vs. charged-particle multiplicity for HF electron decay and theoretical model PYTHIA8.2 \cite{HFe}.}
\end{minipage} 
\end{figure}

Forward-$y$ HFm and midrapidity HFe decay production meaurements as a function of charged-particle multiplicity in pp collisions at $\sqrt{s}$ = 8 and 13 TeV, respectively, are shown in Fig. \ref{data}. A similar faster-than-linear increase is observed and the two measurements are compatible within uncertainties. The HFe measurements are compared with expectations from PYTHIA8.2 event generator \cite{Sj_strand_2015} that are in fair agreement with the data, as shown in Fig. \ref{model}. 

\section{Summary and Outlook}
Measurements of $\textrm{J}/\psi$ , D mesons and electrons from HF hadron decays at midrapidity show a faster-than-linear increase vs. multiplicity with a larger enhancement at high multiplicities. 
No energy dependence was observed in $\textrm{J}/\psi$ measurements at forward rapidity and for midrapidity $\rm J/\psi$ no indication was observed of a possible autocorrelation bias that could arise from the multiplicity selection. Although theoretical models of HF-production including  MPI describe qualitatively the observed trends, more refined theoretical calculations are required to fully understand the evolution from mid- to forward rapidity.\\



\section*{References}
\begin{thebibliography}{9}
\bibitem{charm} ALICE Collaboration, Phys. Rev. D \textbf{105} L011103, (2022)
\bibitem{beauty} ALICE Collaboration, JHEP \textbf{05} 220, (2021)
\bibitem{run1} ALICE Collaboration, Phys.Rev.Lett. \textbf{109} 112301, 2012
\bibitem{run2} ALICE Collaboration, JHEP \textbf{1909} 008, (2019)
\bibitem{run1.1} ALICE Collaboration, Phys.Lett.B \textbf{708} 265-275, (2012)
\bibitem{FONLL} M. Cacciari, M. Greco and P. Nason,  JHEP \textbf{05} 007, (1998)
\bibitem{NNLO} M. Cacciari and M. Greco, Nuclear Physics B \textbf{421} (3) 530-544, (1994)
\bibitem{strangeness} ALICE Collaboration, Nature Physics \textbf{13} 535-539, (2017)
\bibitem{collectivity} ALICE Collaboration, Nucl.Phys.A \textbf{982} 487-490, (2019)
\bibitem{collectivity1} ALICE Collaboration, Phys. Rev. Lett. \textbf{122} 072301, (2019) 
\bibitem{Aamodt}ALICE Collaboration, JINST \textbf{3} S08002, (2008)
\bibitem{dmeson1} ALICE Collaboration, JHEP \textbf{09} 148, (2015)
\bibitem{jpsi} ALICE Collaboration, JHEP \textbf{06} 015, (2022)
\bibitem{HFe} ALICE Collaboration, PoS(\textbf{LHCP2018})044, (2018)
\bibitem{dmeson} ALICE Collaboration, PoS(\textbf{LHCP2021})190, (2021)
\bibitem{Sj_strand_2015} Sj$\ddot{\textrm{o}}$strand Torbj$\ddot{\textrm{o}}$rn, Computer Physics Communications \textbf{191} 159-177, (2015)
\end{thebibliography}
\end{document}